\documentclass[10pt,twocolumn,english]{IEEEtran}

\usepackage[T1]{fontenc}
\usepackage[latin9]{inputenc}
\usepackage{graphicx}
\usepackage{url}
\usepackage{epstopdf}

\makeatletter

\providecommand{\tabularnewline}{\\}

\usepackage{algorithmic}
\usepackage{amsmath}
\usepackage{graphicx}
\usepackage{amssymb}
\usepackage{cite}
\usepackage{array}
\usepackage{multirow}

\makeatother
\usepackage{babel}

\usepackage[backgroundcolor=black!10,bordercolor=white!70!black,linecolor=white!70!black]{todonotes} 
\usepackage{breakcites}
\usepackage{varioref}
\usepackage{xifthen}
\usepackage{amsfonts}
\usepackage{mathtools}
\usepackage{xfrac}
\usepackage{array}
\usepackage{xhfill}
\usepackage{multirow}
\usepackage{booktabs}
\usepackage{arydshln}
\usepackage{listings}
\usepackage{semantic}

\PassOptionsToPackage{hyphens}{url}\usepackage{hyperref} 
\hypersetup{
  pdfauthor={Author},%
  pdfkeywords={password-based authenticated key exchange,formal methods,key agreement},%
  pdftitle={Title},%
  colorlinks,                        
  linkcolor={red!50!black},          
  citecolor={blue!50!black},         
  urlcolor={blue!50!green!90!black}, 
}
\usepackage{hypcap}

\newlength{\protocolArrowLength}
\newcolumntype{C}[1]{>{\centering\let\newline\\\arraybackslash\hspace{0pt}}m{#1}}
\newcommand{\protsend}[1]{\xrightarrow{\makebox[\protocolArrowLength]{$#1$}}}
\newcommand{\protreceive}[1]{\xleftarrow{\makebox[\protocolArrowLength]{$#1$}}}

\newcommand{\inr}{\stackrel{\mathrm{R}}{<-}}
\newcommand{\pbrackets}[1]{\left ( #1 \right )}
\newcommand{\sbrackets}[1]{\left [ #1 \right ]}

\newcommand{\set}[1]{\left \{ #1 \right \}}
\newcommand{\kstar}{^\star}
\newcommand{\typeset}[1]{\mathcal{#1}}

\newcommand{\intset}{\mathbb{Z}}
\newcommand{\groupset}{\mathbb{G}}

\newcommand{\pair}[2]{\left ( #1, #2 \right )}
\newcommand{\of}[1]{\left ( #1 \right )}

\newcommand{\procstop}{\mathbf{0}}
\newcommand{\mn}[1]{\mathrm{#1}}

\newcommand{\piread}[2]{\mn{in}\of{#1, #2}}
\newcommand{\piwrite}[2]{\mn{out}\of{#1, #2}}
\newcommand{\pilet}[2]{\mn{let}\ {#1}\ \mn{in}\ {#2}} 

\newcommand{\piinsert}[2]{\mn{insert}\ #1\of{#2}}
\newcommand{\piget}[3]{\mn{get}\ #1\of{#2}\ \mn{in}\ {#3}} 
\newcommand{\pigetse}[5]{\mn{get}\ #1\of{#2}\ifthenelse{\isempty{#3}}{}{\ \mn{suchthat}\ {#3}}\ \mn{in}\ {#4}\ifthenelse{\isempty{#5}}{}{\ \mn{else}\ {#5}}} 

\newcommand{\resultna}{-}
\newcommand{\resultve}{\checkmark}
\newcommand{\resultfa}{\times}

\begin{document}

\title{Analysing and Patching SPEKE in ISO/IEC} 

\author{Feng Hao,
        Roberto Metere, 
        Siamak F.~Shahandashti and
        Changyu Dong
\thanks{Feng Hao, Roberto Metere and Changyu Dong are with School of Computing, Newcastle University, United Kingdom. 
E-mail: \{feng.hao, r.metere2, changyu.dong\}@ncl.ac.uk. 
Siamak F.~Shahandashti is with the Department of Computer Science, University of York, United Kindom. Email: siamak.shahandashti@york.ac.uk. This work is supported by ERC Starting Grant, No.~306994. Dong is supported by EPSRC EP/M013561/2.}
}

\maketitle

\begin{abstract}

Simple Password Exponential Key Exchange (SPEKE) is a well-known Password Authenticated Key Exchange (PAKE) protocol that has been used in Blackberry phones for secure messaging and Entrust's TruePass end-to-end web products. It has also been included into international standards such as ISO/IEC 11770-4 and IEEE P1363.2. In this paper, we analyse the SPEKE protocol as specified in the ISO/IEC and IEEE standards. We identify that the protocol is vulnerable to two new attacks: an impersonation attack that allows an attacker to impersonate a user without knowing the password by launching two parallel sessions with the victim, and a key-malleability attack that allows a man-in-the-middle (MITM) to manipulate the session key without being detected by the end users. Both attacks have been acknowledged by the technical committee of ISO/IEC SC 27, and ISO/IEC 11770-4 revised as a result. We propose a patched SPEKE called P-SPEKE and present a formal analysis in the Applied Pi Calculus using ProVerif to show that the proposed patch prevents both attacks. The proposed patch has been included into the latest revision of ISO/IEC 11770-4 published in 2017. 


\end{abstract}

\section{Introduction}
\label{sec:introduction}
A password-authenticated key exchange (PAKE) protocol aims to establish a high-entropy session key for secure communication between two parties based on a low-entropy secret password known to both without relying on any external trusted parties. The idea of bootstrapping a high-entropy secret key based on a low-entropy secret password is counter-intuitive, and for a long time had been thought impossible until the seminal work by Bellovin and Merrit who proposed the first PAKE solution called Encrypted Key Exchange (EKE)~\cite{bellovin1992encrypted}. Since then, research on PAKE has become a thriving field: many PAKE protocols have been proposed, and some have been included into international standards~\cite{ieeep1363.2d26,isoiec11770-4:2006}.

However, the original EKE protocol was found to suffer from several limitations, of which the most significant one was the leakage about the password~\cite{jaspan1996dual}. Motivated by addressing the limitations, Jablon proposed another PAKE solution called the simple password exponential key exchange (SPEKE) in 1996~\cite{jablon1996strong}. SPEKE proves to be a more practical protocol than EKE since it does not have the same password leakage problem as in EKE. Although researchers raised concerns on some other aspects of SPEKE~\cite{zhang2004analysis,tang2005security} such as the possibility for an online attacker to test multiple passwords in one go, no major flaws have been reported. Over the years, SPEKE has been used in several commercial applications: for example, the secure messaging on Blackberry phones~\cite{blackberry2016security} and Entrust's TruePass end-to-end web products~\cite{entrust2003entrust}. SPEKE has also been included into the international standards such as IEEE P1363.2~\cite{ieeep1363.2d26} and ISO/IEC 11770-4~\cite{isoiec11770p4}. 

Given the wide usage of SPEKE in practical applications and its inclusion in standards, we believe a thorough analysis of SPEKE is both necessary and important. In this paper, we revisit SPEKE and its variants specified in the original paper~\cite{jablon1996strong}, the IEEE 1363.2~\cite{ieeep1363.2d26} and ISO/IEC 11770-4~\cite{isoiec11770-4:2006} standards. We first observe that the original SPEKE protocol is subtly different from those defined in the standards. The difference  has significant security implications, which are not explained in the standards. 

During the review, we have identified several issues that have not been reported before. In particular, we find two new attacks on SPEKE: namely, an impersonation attack and a key-malleability attack. The first attack allows an attacker to impersonate a user without knowing the password by launching two parallel sessions with the victim. The second attack allows an attacker to manipulate the session key without being detected. To address the identified problems, we propose a patched SPEKE, called P-SPEKE, which prevents both attacks by including the user identities in the key derivation function without altering the symmetry of the original SPEKE protocol. We build a formal model in the Applied Pi Calculus using ProVerif and apply it to formally analyse P-SPEKE. Our analysis confirms that the proposed patch is immune to the attacks. Finally, we identify an efficiency problem with the key confirmation procedure specified in both the ISO/IEC and IEEE standards and accordingly propose an improved procedure. 

Our contributions are summarized below.


\begin{itemize}
\item We discover two new attacks on SPEKE: an impersonation attack and a key-malleability attack. We explain how the attacks affect the SPEKE variants specified in the IEEE P1363.2 and ISO/IEC 11770-4 standards. 

\item We propose a patched SPEKE, called P-SPEKE, which prevents both attacks without altering the symmetry of the SPEKE protocol. Furthermore, we propose an improved key confirmation procedure, which is more round-efficient than the one defined in the standards. 

\item We build a formal model in the Applied Pi Calculus and verify the proposed patch by using ProVerif. Our formal analysis confirms that the proposed patch is immune against the identified attacks. 
\end{itemize}

This paper extends the earlier conference paper~\cite{hao2014speke} by adding a formal analysis of the patched SPEKE protocol, and details of how the proposed patch was accepted and included into the revision of ISO/IEC 11770-4. The two attacks and the efficiency issues, initially reported in~\cite{hao2014speke}, were discussed and acknowledged by the technical committee of ISO/IEC SC 27, Working Group 2. Accordingly, the ISO/IEC 11770-4 standard was revised. The latest revision ISO/IEC 11770-4:2017, incorporating our proposed patch and the improved key confirmation procedure, was formally published in November 2017~\cite{isoiec11770p4}.

\section{Background}\label{sec:background}

\subsection{Password Authenticated Key Exchange}

Since the invention of the first PAKE solution in~\cite{bellovin1992encrypted}, many PAKE protocols have been proposed, among which only a few have been actually used in practice. Notable examples of PAKE that have been deployed in practical applications include EKE~\cite{bellovin1992encrypted}, SPEKE~\cite{jablon1996strong} and J-PAKE~\cite{hao2008password}. These three protocols happen to represent three different ways of constructing a PAKE. EKE works by using the shared password as a symmetric key to encrypt Diffie-Hellman key exchange items. Variants of EKE (e.g., SPAKE2~\cite{abdalla2005simple}) often differ only in how the symmetric cipher is instantiated. SPEKE works by using the shared password to derive a secret group generator for performing Diffie-Hellman key exchange. There are variants of SPEKE, such as Dragonfly~\cite{harkins2008simultaneous} and PACE~\cite{bender2012pace}, which use different methods to derive the secret generator from the password. J-PAKE works by using the password to randomize the secret exponents in order to achieve a cancellation effect. A distinctive feature of J-PAKE as compared to the other two is its use of Zero Knowledge Proof (ZKP) to enforce participants to follow the protocol specification. By comparison, the use of ZKP is considered incompatible with the design of EKE and SPEKE. 

A PAKE protocol serves to provide two functions: authentication and key exchange. The former is based on the knowledge of a password. If the two passwords match at both ends, a session key will be created for the subsequent secure communication. In the following, we review some common properties of a secure password authenticated key exchange protocols based on~\cite{bellovin1992encrypted, jablon1996strong, hao2008password}; we also refer the reader to classic definitions of authentication from Lowe~\cite{lowe1997hierarchy}. Formal treatments of PAKE, based on authenticated key exchange models proposed by Bellare and Rogaway in 1993~\cite{bellare1993entity}, can be found in~\cite{bellare2000authenticated,goldreich2001session,katz2001efficient,abdalla2015security}.

\noindent\textbf{Correctness}. 
In the setting of key-exchange protocols, the protocol is correct if it gives both authentication and key distribution in presence of honest parties~\cite{woo1993semantic}. This is a basic and necessary step in a formal model to prove that without influence of attackers, honest parties should always complete the protocol as expected.

\noindent\textbf{Secrecy of the pre-shared password}. 
This property requires that the execution of the protocol must not reveal
any data that would allow an attacker to learn the password through off-line exhaustive search. 
If the attacker is directly engaging in the key exchange, he should be limited to guess only one password per protocol execution. 

\noindent\textbf{Implicit key authentication}. 
Assume the key exchange protocol is run between Alice and Bob. The protocol is said to provide implicit key authentication if Alice is assured that no one other than Bob can compute the session key~\cite{stinson2005cryptography}.

\noindent\textbf{Explicit key authentication}. 
Explicit authentication can only be achieved with a confirmation phase~\cite{stinson2005cryptography}.
This property requires that the entities have actually computed the {\em same} key.
It completes and strengthens the implicit key authentication; in fact, if the two participants are the sole entities who can learn the session key {\em and} they have actually computed the key, the successive communication shall be secure.

\noindent\textbf{Weak and strong entity authentication}. 
{\em Weak} or {\em strong} entity authentication respectively correspond to the {\em weak agreement} and {\em injective agreement} properties of Lowe~\cite{lowe1997hierarchy}.
A protocol achieves weak authentication if a participant believes she is speaking with another participant, and the other participant indeed started an authentication process with her.
Even though this may seem a sufficient property for mutual authentication, it is not.
In fact, nothing can be said about the problem where the party is tricked to communicate with some replayed session of the other party.
With strong authentication, the additional property of agreeing with both the session and the session key is required.
Strong entity authentication ensures that replay attacks and man-in-the-middle attacks are prevented.

\noindent\textbf{Perfect forward secrecy}.
Perfect forward secrecy (PFS) ensures that the confidentiality of past session keys is preserved even when the long term secret, i.e., the password, is disclosed. This property implies that an attacker who knows the password still cannot learn the session key if he only \emph{passively} eavesdrops the key exchange process.

\subsection{The original SPEKE}
\label{sec:original-speke}
The original specification of the SPEKE protocol in Jablon's 1996 paper~\cite{jablon1996strong} is as follows.
Participants agreed on a group $\groupset$ of safe prime order $p = 2q+1$ where $q$ is also a prime. The SPEKE protocol operates in the subgroup of $\groupset$ of prime order $q$ where the discrete logarithm problem is assumed to be hard. Two remote parties, Alice and Bob, share a common secret password $s$ from which they apply a function $f(\cdot)$ to compute the group generator: $g = f(s) = s^2 \bmod{p}$.
Unless specified otherwise, all modular operations in the rest of the paper are performed with respect to the modulus $p$. We will omit ``$\bmod{\;p}$'' in the notation for simplicity.

\begin{figure}[htbp]
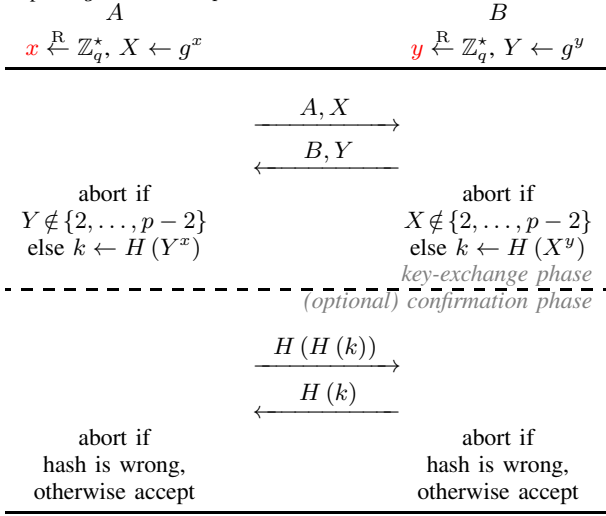

\small
\centering
\caption{Original SPEKE scheme. $A$ and $B$ share the password $s$ and computed $g = s^2 \mod p$.}
\label{fig:speke_jablon}
  \setlength{\protocolArrowLength}{50pt}
  \begin{centering}
    \begin{tabular}{cC{\protocolArrowLength}c}
    $A$ & & $B$ \\
    ${\color{red}x} \inr \intset\kstar_q$, $X <- g^x$ &  & ${\color{red}y} \inr \intset\kstar_q$, $Y <- g^y$ \\ \midrule
      & $\begin{array}{c}
          \protsend{A, X} \\
          \protreceive{B, Y}
        \end{array}$ & \\
    abort if & & abort if \\
    $Y\!\notin\!\set{2,\dots, p-2}$ & & $X\!\notin\!\set{2, \dots, p-2}$ \\
    else $k <- H\of{Y^x}$ & & else $k <- H\of{X^y}$ \\
      \multicolumn{3}{r}{\color{gray} \em key-exchange phase} \\ \cdashline{1-3}
      \multicolumn{3}{r}{\color{gray} \em (optional) confirmation phase} \\
      & $\begin{array}{c}
          \protsend{H\of{H\of{k}}} \\
          \protreceive{H\of{k}}
        \end{array}$ & \\
    abort if & & abort if \\
    hash is wrong, & & hash is wrong, \\
    otherwise accept & & otherwise accept \\
    \bottomrule
    \end{tabular}
  \end{centering}
\end{figure}

The SPEKE protocol runs in two phases: the {\em key-exchange phase} and the {\em key-confirmation phase}, as illustrated in Figure~\ref{fig:speke_jablon}. 
In the first phase, Alice chooses a secret value $x$ uniformly at random in $\intset\kstar_q = \set{1, \dots, q-1}$, and sends $g^x$ to Bob. Similarly, Bob chooses a secret value $y$ uniformly at random in $\intset\kstar_q$, and sends $g^y$ to Alice. Upon receiving $g^y$, Alice verifies that its value is between $2$ and $p-2$. This is to prevent the small subgroup confinement attack. Subsequently, Alice computes a session key $k = H((g^y)^x) = H(g^{xy})$ where $H$ is a cryptographic hash function (used as a key derivation function here). Similarly Bob verifies $g^x$ belongs to $\set{2, \dots, p-2}$ and then computes the same session key $k = H((g^x)^y) = H(g^{xy})$. The key-exchange phase is completely symmetric. The symmetry in the design helps simplify the security analysis and reduce the communication rounds especially in a mesh network. 

The second phase serves to provide explicit assurance that both parties have actually derived the same session key. This is realized in the original SPEKE paper~\cite{jablon1996strong} as follows: one party sends $H(H(k))$ first and the other party replies with $H(k)$ later. 

The above key confirmation method has two subtle issues. First, it is ambiguous which party should send $H(H(k))$ first. As we will explain, this ambiguity also carries over to the SPEKE specifications in the ISO/IEC and IEEE standards. Second, from a theoretical perspective, the direct use of the session key in the key confirmation process renders the session key no longer indistinguishable from random after the key confirmation is finished, hence breaking the session-key indistinguishability requirement in a formal model~\cite{bellare1993entity}. 

In the standards, the key confirmation phase is optional and it is left to the applications to decide whether it is added.  With the absence of this phase, key confirmation will have to be deferred to the later secure communication stage where the session key is used to encrypt and decrypt messages (in the authenticated mode) and the decryption will only work if the session keys used at the two sides are equal. 

\subsection{Previous attacks}
\label{sec:previous-attacks}
In~\cite{zhang2004analysis}, Zhang proposed an exponential-equivalence attack on SPEKE. This attack exploits the fact that some passwords are exponentially related. For example, two different passwords $s$ and $s'$ may have the relation that $s' = s^r \bmod{p}$ where $r$ is an arbitrary integer ($r \neq 1$). By exploiting this relation, an active attacker can rule out two passwords in one go, and in the general case can rule out multiple passwords in one go if they are all exponentially related. This attack is especially problematic when the password is digits-only, e.g., a Personal Identification Numbers (PIN). As a countermeasure, Zhang proposed  to hash the password before taking the square operation: in other words, redefining the password mapping function to $f(s) = (H(s))^2 \bmod{p}$. The hashing of passwords makes it much harder for the attacker to find exponential equivalence among the hashed outputs. Zhang's attack is acknowledged in IEEE P1363.2~\cite{ieeep1363.2d26}, which adds a hash function in SPEKE when deriving the base generator from the password.

Tang and Mitchell illustrated three attacks on the SPEKE protocol~\cite{tang2005security}.
The first attack is essentially the same as Zhang's~\cite{zhang2004analysis}: an active attacker is able to test multiple passwords in one execution of the protocol by exploiting the exponential equivalence of passwords. The authors suggest to hash the identities of the parties along with the password to get the generator, that is $g = H\of{s\|A\|B}$ where $A$ and $B$ are identities of two communicating parties. However, this countermeasure has the limitation that it breaks the symmetry of the protocol; instead of allowing the two parties to exchange messages simultaneously in one round, the two parties must first agree whose identity should be put first in the hash, which requires extra communication. The second attack is a {\em unilateral} Unknown Key-Share (UKS) attack. In this attack, the user is assumed to share the same password with more than one servers\footnote{We remark that it is unusual to assume a user shares the same password with multiple server in the security model for PAKE, as a server will be able to trivially impersonate another server. However, in practice, many users do reuse passwords across several accounts.}. By replaying messages, the attacker may trick the user into believing that he is sharing a key with one server, but in fact he is sharing a key with a different server. To address the attack, they propose to include the server's identity into the computation of $g$. However, same as before, this countermeasure breaks the symmetry of the original protocol. The last attack they show is a scenario where two sessions are swapped. Here, the two parties run two concurrent sessions, and the attacker swaps the messages between the two sessions.
At the end of the protocol, the parties will have shared two session keys, but they may get confused which message belongs to which session.
They call this a generic vulnerability, which in this paper we call a {\em sessions swap} attack. To address this problem, they propose to include the ``session identifier'' into the computation of $g$, but the paper gives no details on the definition of the ``session identifier''.

\subsection{Specification in standards}
\label{sec:specification-in-standards}

When SPEKE was included into the IEEE P1363.2 and ISO/IEC 11770-4 standards, the protocol was revised to prevent the exponential-equivalence attack reported in~\cite{zhang2004analysis} and~\cite{tang2005security}. In the revised protocol, the password is hashed first before computing a secret generator. More specifically, the generator is obtained from $g = (H(s))^2 \bmod{p}$ instead of $g = s^2 \bmod{p}$ as in the original 1996 paper.

It is also worth noting that the key confirmation procedure of SPEKE defined in the standards is also different from that in the original SPEKE paper~\cite{jablon1996strong}. In IEEE P1363.2~\cite{ieeep1363.2d26} and in ISO/IEC 11770-4:2006~\cite{isoiec11770-4:2006}, the key confirmation is defined as follows.
\begin{equation}
\label{eq:double-hash-key-confirmation}
  \begin{array}{ccccc}
    \mathrm{Alice} & \rightarrow & \mathrm{Bob} & : & H(3 \| g^x \| g^y \| g^{yx} \| g) \\
    \mathrm{Bob} & \rightarrow & \mathrm{Alice} & : & H(4 \| g^x \| g^y \| g^{xy} \| g) \\
  \end{array}
\end{equation}
As explicitly stated in the ISO/IEC 11770-4 standard, there is no order in the above two steps. In the same standard, it is also stated that there is no order during the SPEKE exchange phase. We find the two statements contradictory: the fact that $g^x$ comes before $g^y$ in the definition of key confirmation implies there is an order during the key exchange phase.

We would like to highlight that the above issue was carried over from Jablon's original 1996 paper~\cite{jablon1996strong}, which specifies that ``Alice'' sends the first confirmation message $H(H(k))$. Given the symmetric nature of the protocol, it is ambiguous which party is ``Alice''. This ambiguity was unquestioned at the time of standardization and consequently was inherited by the specifications in IEEE P1363.2 and ISO/IEC 11770-4:2006. 

We presented the above issue to the ISO/IEC SC 27 technical committee. The issue was acknowledged and rectified in the latest revision ISO/IEC 11770-4:2017. We will explain the details of the change later.

\section{New attacks}
\label{sec:new-attacks}

In this section, we present two new attacks that are not reported before: an impersonation attack and a key-malleability attack. We will first explain how the attacks work on the original SPEKE protocol~\cite{jablon1996strong} and then explain their applicability to the SPEKE variants defined in the IEEE and ISO/IEC standards~\cite{ieeep1363.2d26,isoiec11770-4:2006}.

\subsection{Impersonation attack}
\label{sec:impersonation-attack}

The first attack happens in the setting of parallel sessions: a user is engaged with another user in multiple sessions running in parallel. We illustrate the attack of Mallory who will be able to impersonate the user Bob to Alice, by launching parallel sessions with Alice to make Alice believe she is communicating with Bob, but actually Bob is not involved at all in the communication. 

\begin{figure}
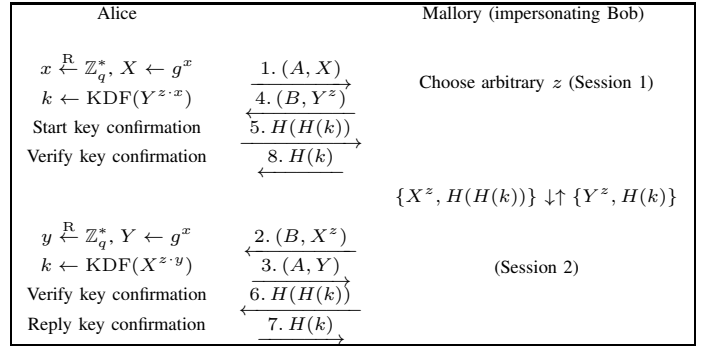

  \scriptsize
  \caption{Impersonation attack on SPEKE}
  \label{fig:impersonation-attack}
  \begin{center}
    \begin{tabular}{|ccc|}
      \hline 
      Alice &  & Mallory (impersonating Bob) \tabularnewline[\doublerulesep]
      \multicolumn{1}{|c}{} &  & \multirow{5}{*}{Choose arbitrary $z$ (Session 1)}\tabularnewline
      $x \inr \mathbb{Z}_{q}^{*}$, $X\leftarrow g^{x}$ & $\underrightarrow{\,\,\,1.\,(A,X)\,\,\,}$ & \tabularnewline
      $k\leftarrow\mathrm{KDF}(Y^{z\cdot x})$ & $\underleftarrow{\,\,\,4.\,(B,Y^{z})\,\,\,}$ & \tabularnewline
      Start key confirmation & $\underrightarrow{\,\,\,5.\,H(H(k))\,\,\,}$ & \tabularnewline
      Verify key confirmation & $\underleftarrow{\,\,\,8.\,H(k)\,\,\,}$ & \tabularnewline
      &  & \multirow{2}{*}{$\{X^{z},H(H(k))\}\downarrow\uparrow\{Y^{z},H(k)\}$}\tabularnewline
      &  & \tabularnewline
      $y \inr \mathbb{Z}_{q}^{*}$, $Y\leftarrow g^{x}$ & $\underleftarrow{\,\,\,2.\,(B,X^{z})\,\,\,}$ & \multirow{4}{*}{(Session 2)}\tabularnewline
      $k\leftarrow\mathrm{KDF}(X^{z\cdot y})$ & $\underrightarrow{\,\,\,3.\,(A,Y)\,\,\,}$ & \tabularnewline
      Verify key confirmation & $\underleftarrow{\,\,\,6.\,H(H(k))\,\,\,}$ & \tabularnewline
      Reply key confirmation & $\underrightarrow{\,\,\,7.\,H(k)\,\,\,}$ & \tabularnewline
      \hline 
    \end{tabular}
  \end{center}
\end{figure}

The attack is illustrated in Figure~\ref{fig:impersonation-attack}. Details of each step are explained below.

\begin{enumerate}

\item Alice chooses a secret exponent $x$ and computes $X \leftarrow g^x$. She initiates the protocol by sending $A, X$ to the insecure channel.

\item Mallory is in control of the channel and intercepts all the messages to Bob who never receives anything. So, Mallory receives the first message from Alice and generates an exponent $z$ such that $X^z \in \set{2, \dots, p - 2}$\footnote{When $z = 1$ the work of Mallory reduces to simply relaying Alice's messages to herself in the other session, which may be detected if Alice checks for duplicate of messages.}.
Mallory, impersonating Bob, initiates a parallel SPEKE session with Alice by sending her $B, X^z$.

\item Alice follows the second session generating an exponent $y$ and computing $Y \leftarrow g^y$. She sends $A, Y$ to the insecure channel.

\item Mallory intercepts the message and raises it to the power of $z$ (with overwhelming probability, $Y^z$ will not be $1$ or $p-1$). Then, Mallory sends back to Alice $B, Y^z$ in the first session.

\item At this point, Alice computes the key $k = H\of{(Y^z)^x} = H\of{g^{xyz}}$ for the first session, generates the key confirmation challenge $H(H(k))$, and sends it to Bob.

\item Mallory intercepts the challenge from the first session and relays it to Alice in the second session.

\item Following the protocol, Alice answers the challenge with $H\of{k}$.

\item Finally, Mallory intercepts Alice's answer in the second session and replays it in the first session to pass the key confirmation procedure. 

\end{enumerate}

At the end of the above attack, Alice authenticates Mallory as ``Bob'' in both sessions. However, Mallory does not know any secret password and the real ``Bob'' has never been involved in this key exchange. This indicates a serious flaw in the authentication procedure. We should note that in the above attack, we assume the initiator of the session is responsible for sending the first key confirmation message. This is allowed by the protocol since SPEKE specifications in both the IEEE and ISO/IEC standards permit the two parties to start the key confirmation in any order. 

This attack can be regarded as a special instance of the Unknown Key-Share (UKS) attack~\cite{hao2010robust}. Alice thinks she is communicating with ``Bob'', but actually she is communicating with another instance of herself. This confusion of identities in the key establishment can cause problems in some scenarios.
For example, using the derived session key $k$ in an authenticated mode, like AES-GCM, Alice may send an encrypted message to Bob: ``Please pay Charlie 5 bitcoins''.
Mallory can intercept this message and (without knowing its content) relay back to Alice in the second session.
Since the message is verified to be authentic from ``Bob'', Alice may follow the instruction (assume Alice is an automated program that follows the protocol).
Thus, although Alice's initial intention is to make Bob pay Charlie 5 bitcoins, she ends up paying Charlie instead.

\subsection{Key-malleability attack}
\label{sec:malleability-attack}

\begin{figure*}
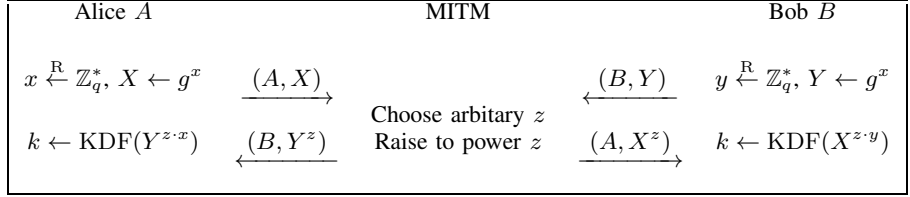

\small
  \caption{Key-malleability attack on SPEKE}
  \label{fig:key-malleability-attack}
  \begin{center}
    \begin{tabular}{|ccccc|}
      \hline 
      Alice $A$ &  & MITM &  & \multicolumn{1}{c|}{Bob $B$}\tabularnewline[\doublerulesep]
      \multicolumn{1}{|c}{} &  &  &  & \tabularnewline
      $x \inr \mathbb{Z}_{q}^{*}$, $X\leftarrow g^{x}$ & $\underrightarrow{\,\,\,(A,X)\,\,\,}$ &  & $\underleftarrow{\,\,\,\,(B,Y)\,\,\,}$ & $y \inr \mathbb{Z}_{q}^{*}$, $Y\leftarrow g^{x}$\tabularnewline
      &  & Choose arbitary $z$ &  & \tabularnewline
      $k\leftarrow\mathrm{KDF}(Y^{z\cdot x})$ & $\underleftarrow{\,\,\,(B,Y^{z})\,\,\,}$ & Raise to power $z$ & $\underrightarrow{\,\,\,(A,X^{z})\,\,\,}$ & $k\leftarrow\mathrm{KDF}(X^{z\cdot y})$\tabularnewline
      &  &  &  & \tabularnewline
      \hline 
    \end{tabular}
  \end{center}
\end{figure*}

The second attack is a man-in-the-middle attack as shown in Figure~\ref{fig:key-malleability-attack}. The attacker chooses an arbitrary $z$ from $\set{2, \dots, q-2}$, raises the intercepted item to the power of $z$ and passes it on. The parties at the two ends are still able to derive the same session key $k = H(g^{xyz})$, but without being aware that the messages have been modified.

The fact that an attacker is able to manipulate the session key without being detected has significant implications on the theoretical analysis of the protocol. In the original SPEKE paper, the protocol has no security proofs; it is heuristically argued that the security of the session key in SPEKE depends on either the Computational Diffie-Hellman assumption (i.e., an attacker is unable to compute the session key) or the Decisional Diffie-Hellman assumption (i.e., an attacker is unable to distinguish the session key from random). The existence of such a key-malleability attack suggests that a clean reduction to CDH or DDH is not possible. As an example, $z$ can be a result of an arbitrary function $f(\cdot)$ with the incepted inputs, i.e., $z = f(g^x, g^y)$. Because of the correlation of values on the exponent, standard CHD and DDH assumptions are not applicable since they require the secret values on the exponent be \emph{independent}.


\subsection{Discussion on standards}

\subsubsection{Explicit key confirmation} 

Recall from Section~\ref{sec:specification-in-standards} that the SPEKE schemes specified in the standards differ from the original SPEKE paper in how the explicit key confirmation is defined. More specifically, the key confirmation procedure in IEEE P1363.2 and ISO/IEC 11770-4 includes additional data in the hash: i.e., key exchange items $g^x$ and $g^y$. This change does not prevent the impersonation attack; the attacker is still able to relay the key confirmation string in one session to another parallel session to accomplish mutual authentication in both sessions. However, the key-malleability attack no longer works if the key confirmation method in IEEE 1363.2 or ISO/IEC 11770-4 is used. We should emphasize that the key confirmation method in both standards are marked as ``optional''. Hence, the key-malleability attack is still applicable to the \emph{implicitly authenticated} version of the SPEKE in both standards.

\subsubsection{Definition of shared secret}  

In the earlier conference version of the paper~\cite{hao2014speke}, we point out that the definition of the shared secret in ISO/IEC 11770-4:2006~\cite{isoiec11770-4:2006} is ambiguous. The shared low-entropy secret, denoted $\pi$ in that standard document~\cite{isoiec11770-4:2006}, is defined as follows.

\begin{quote}
	\emph{``A password-based octet string which is generally derived from a password or a hashed password, identifiers for one or more entities, an identifier of a communication session if more than one session might execute concurrently, and optionally includes a salt value and/or other data.}
\end{quote}

The above definition seems to include the ``identifiers for one or more entities'' as part of the shared secret. If the entity identifiers were included, the impersonation attack would not work, but the key-malleability would still work. However, the standard does not provide any formula about $\pi$. It is not even clear if one or both entities' identifiers should be included, and if only one identifier is to be included, which one and how. Furthermore, the word ``generally'' weakens the rigour in the definition and makes it subject to potentially different interpretations. 

By comparison, the definition of the shared secret in IEEE P1363.2 (D26)~\cite{ieeep1363.2d26} is clearer. It is specified as follows:

\begin{quote}
	\emph{``A password-based octet string, used for authentication. $\pi$ is generally derived from a password or a hashed password, and may incorporate a salt value, identifiers for one or more parties, and/or other shared data.''}
\end{quote}

This definition clearly indicates that the incorporation of ``a salt value, identifiers for one or more parties, and/or other shared data'' is not mandatory (as indicated by the use of the word ``may''). Based on the definition, it is clear that both attacks are applicable to the SPEKE scheme defined in IEEE P1363.2.


The issue about the ambiguity in the definition was acknowledge by ISO/IEC SC 27 after we first pointed it out in~\cite{hao2014speke}, and was rectified accordingly. In the latest revision in ISO/IEC 11770-4:2017, the definition of the shared secret has been revised to follow the same as in IEEE P1363.2 (D26)~\cite{ieeep1363.2d26}. In this revision, the two reported attacks are addressed by making technical changes to the SPEKE specification, as we will explain in the next section.

\section{Solution}

\subsection{Patched SPEKE}

There are several reasons to explain the cause of the two attacks. First, there is
no reliable method in SPEKE to prevent a sent message being relayed back to
the sender. Second, there is no mechanism in the protocol to verify the integrity
of the message, i.e., whether they have been altered during the transit. Third,
no user identifiers are included in the key exchange process. It may be argued
that all these issues can be addressed by using a Zero Knowledge Proof (ZKP)
(as done in~\cite{hao2008password}). However, in SPEKE, the generator is a secret, which makes
it incompatible with any existing ZKP construction. Since the use of ZKP is
impossible in SPEKE, we need to address the attacks in a different way.

Our proposed patch is to redefine the session key computation. Assume Alice sends $g^x$ and Bob sends $g^y$. The session key computation is defined below. 
\begin{align}
  s_A &=  H (A \| g^x) \nonumber \\
  s_B &=  H (B \| g^y) \nonumber \\
  sID &=  \mathrm{max}(s_A, s_B) \| \mathrm{min}(s_A, s_B) \nonumber \\
  k &=  \mathrm{KDF}(sID \| g^{xy}) \label{eq:new-session-key}
\end{align}
When the two users are engaged in multiple concurrent sessions, they need to ensure the identifiers are unique between these sessions. As an example, assume Alice and Bob launch several concurrent sessions. They may use ``Alice'' and ``Bob'' in the first session. When launching a second concurrent session, they should add an extension to make the entity identifier unique -- for example, the entity identifiers may become ``Alice (2)'' and ``Bob (2)'' respectively in the second session, and so on. The use of the extension is to make the entity identifier distinguishable among multiple sessions running in parallel.

The new definition of the session-key computation function in Eq.~\ref{eq:new-session-key} prevents both the impersonation and key-malleability attacks (as well as the \emph{session swap} attack reported in ~\cite{tang2005security}), which we will formally prove in the next section. The key confirmation remains ``optional" as it is currently defined in the  standards.
Furthermore, this patch preserves the optimal one-round efficiency of
the original SPEKE protocol. 

An alternative patch, suggested in the earlier conference paper~\cite{hao2014speke}, is to refine the session key computation as follows.
  \begin{align}
    M &= H\left(\mathrm{min}(A, B) \| \mathrm{max}(A, B)\right) \nonumber \\
    N &= H\left(\mathrm{min}(g^x, g^y) \| \mathrm{max}(g^x, g^y)\right) \nonumber \\
    k &= \mathrm{KDF}(M, N, g^{xy})
  \end{align}

\begin{figure*}
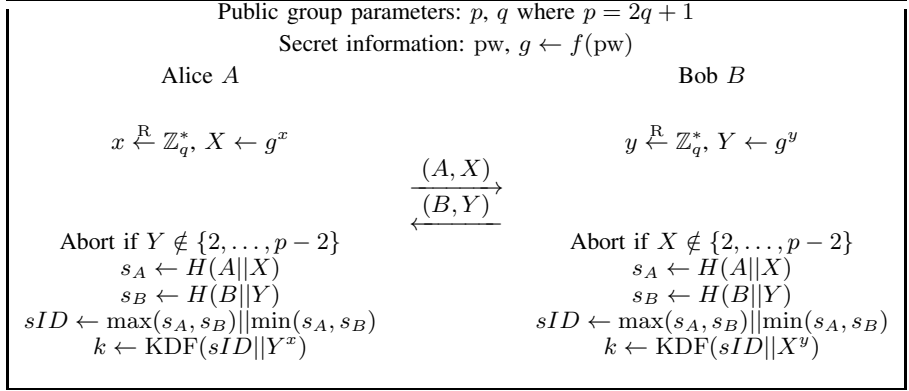

\small
  \caption{Patched SPEKE (included in ISO/IEC 11770-4:2017~\cite{isoiec11770p4})}
  \label{fig:Patched-SPEKE}
  \begin{center}
    \begin{tabular}{|ccc|}
      \hline 
      \multicolumn{3}{|c|}{Public group parameters: $p$, $q$ where $p=2q+1$}\tabularnewline[\doublerulesep]
      \multicolumn{3}{|c|}{Secret information: $\mathrm{pw}$, $g\leftarrow f(\mathrm{pw})$}\tabularnewline[\doublerulesep]
      Alice $A$ &   & Bob $B$\tabularnewline[\doublerulesep]
      \multicolumn{1}{|c}{} &  & \tabularnewline
      $x \inr \mathbb{Z}_{q}^{*}$, $X\leftarrow g^{x}$ &  & $y \inr \mathbb{Z}_{q}^{*}$, $Y\leftarrow g^{y}$\tabularnewline
        & $\underrightarrow{\,\,\,(A,X)\,\,\,}$ & \tabularnewline
        & $\underleftarrow{\,\,\,(B,Y)\,\,\,}$ & \tabularnewline
      Abort if $Y \notin \set{2,\dots,p-2}$ &   & Abort if $X \notin \set{2,\dots,p-2}$\tabularnewline
      $s_{A}\leftarrow H(A||X)$ &  & $s_{A}\leftarrow H(A||X)$\tabularnewline
      $s_{B}\leftarrow H(B||Y)$ &  & $s_{B}\leftarrow H(B||Y)$\tabularnewline
      $sID\leftarrow\mathrm{max}(s_{A},s_{B})||\mathrm{min}(s_{A},s_{B})$ &  & $sID\leftarrow\mathrm{max}(s_{A},s_{B})||\mathrm{min}(s_{A},s_{B})$\tabularnewline
      $k\leftarrow\mathrm{KDF}(sID||Y^{x})$ &  & $k\leftarrow\mathrm{KDF}(sID||X^{y})$\tabularnewline
      &  & \tabularnewline
      \hline 
    \end{tabular}
  \end{center}
\end{figure*}

As we will formally analyze in Section~\ref{sec:formalisation}, the above solution also prevents the two attacks. However, the advantage of the solution in Eq.~\ref{eq:new-session-key} is that the hash output has a fixed bit length, which makes it easier to implement the $\mathrm{max}$ and $\mathrm{min}$ function. The final patch, which has been included into the latest revision of ISO/IEC 11770-4 published in 2017, is summarized in Figure~\ref{fig:Patched-SPEKE}.

\subsection{Improved key confirmation}

As highlighted in Section~\ref{sec:specification-in-standards}, neither of the key confirmation procedures defined in IEEE P1363.2 (D26) and ISO/IEC 11770-4 (2006) is symmetric. In both standards, they state that there is ``no special ordering'' of the key confirmation message. This implies that the messages can be sent simultaneously within one round. But in fact, these procedures require two rounds instead of one, because the second message depends on the first. This issue also applies to the key confirmation method in Jablon's original 1996 paper~\cite{jablon1996strong}. If both parties attempt to send the first message at the same time without an agreed order, they cannot tell if the message that they receive is a genuine challenge or a replayed message, and consequently enter a deadlock.

To address the above issue, we propose an improved key confirmation method which preserves the symmetry of the protocol and hence allows the key confirmation to be completed within one round. It works as follows.
\begin{equation}
  \label{eq:new-key-confirm}
  \begin{array}{ccccc}
  \mathrm{Alice} & \rightarrow & \mathrm{Bob} & : &  H(A \| B \| g^x \| g^y \|g^{xy} \| g) \\
  \mathrm{Bob} & \rightarrow & \mathrm{Alice} & : & H(B \| A \| g^y \| g^x \| g^{xy} \| g)
  \end{array}
\end{equation}

An alternative solution, proposed in our earlier paper~\cite{hao2014speke}, is based on NIST SP 800-56A Revision 1 \cite{barker2013recommendation}. It works as follows.
\begin{equation*}
\begin{array}{ccc}
\mathrm{Alice} \rightarrow \mathrm{Bob} & : &  \mathrm{MAC}(k_c, \mathrm{``KC\_1\_U"} \| A \| B \| g^x \| g^y) \\
\mathrm{Bob} \; \rightarrow \mathrm{Alice} & : & \mathrm{MAC}(k_c , \mathrm{``KC\_1\_U"}, \| B \| A \| g^y \| g^x)
\end{array}
\end{equation*}
In the above method, $\mathrm{MAC}$ is a message authenticated code (MAC) algorithm, the string ``KC\_1\_U'' refers to unilateral key confirmation, and $k_c$ is a MAC key. To allow the session key to remain indistinguishable from random even after the key confirmation phase, $k_c$ should be derived differently from the session key, e.g., by adding a specific parameter to the key derivation function say $k_c = \mathrm{KDF}(g^{xy}, \mathrm{``KC"})$. There is no dependence between the two flows, so Alice and Bob can send messages in one round. 
During the revision of ISO/IEC 11770-4, the hash based key confirmation method in Eq.~\ref{eq:new-key-confirm} was preferred and was included into the latest standard since it requires minimum changes in the standard.

\section{Formal analysis}
\label{sec:formalisation}

In this section, we build a formal model using the Applied Pi Calculus, then we apply this model to formally analyse the proposed patch in comparison to existing SPEKE variants.

\subsection{Reasoning in the Applied Pi Calculus in ProVerif}
\label{sec:intro-proverif}
ProVerif~\cite{blanchet2001efficient} is a tool for reasoning in the symbolic model. It has  proved successful in formally verifying dozens of protocols, and has been widely accepted by the community.
ProVerif's input language is a dialect of the Applied Pi Calculus~\cite{blanchet2016modeling,abadi2016applied}, and we limit our informal illustration of the language to the subset that will be useful for describing our model.

The language is strongly typed, and arbitrary types can be declared.
We use the following types for our model: $\typeset{H}$ for the hosts, $\typeset{S}$ for session IDs, $\typeset{K}$ for session keys, then we have the order-$q$ sub-group of $\intset\kstar_p$ where $p=2q+1$ is a safe prime, and the elements in this sub-group can be expressed as $g^x$ where $g$ is a generator and $x$ is from $\intset\kstar_q$.

The main abstraction of the Pi Calculus is the process.
A process describes the algorithm that an entity follows according to the specifications of a protocol scheme.
They can (i) include variables, constants, functions, and (private) nonces (i.e. $\nu x. P$ restricts the value $x$ to the process $P$) (ii) write to and read from any channel $c$, denoted by $\piwrite{c}{\_}$ and $\piread{c}{\_}$ respectively, (iii) insert and extract elements to and from any table $t$, $\piinsert{t}{\_}$ and $get\ t(\_)$, and (iv) record events.
Processes can be put in sequential or parallel execution with other processes including themselves, for unbounded number of times of replication.
Such instruments allow for symbolic modelling of protocols.

To model security properties, the language offers some facility.
The secrecy of names is verified in terms of unreachability and indistinguishability.
The unreachability of the secret by the attacker determines whether the knowledge of the attacker can be augmented with such secret by using the inference rules determined with respect to the model.
From the point of view of indistinguishability, the tool determines whether the attacker can distinguish between executions that use different secrets.

More sophisticated security properties, like entity authentication, bilateral unknown shared-key resilience, and others, can be formalised through events and correspondences~\cite{blanchet2009automatic}.
Events must be explicitly included as extra lines into the processes, can take arguments, and will be recorded in the traces of execution.
Correspondences are implications related to execution of events.
By default the content of events is not accessible to the attacker, until the attacker is already aware of them or it will be by other rules.
Moreover, the attacker is not directly capable of recording events, but it may induce processes to do so.
Loosely speaking, an event can be thought as a piece of meta-semantics with respect to the purpose of the process itself.
For example, the event $e\of{A,B}$ may be interpreted as ``Alice believes of having started an authentication with Bob''.
Although the terminology ``Alice believes\dots'' can recall the BAN logic~\cite{abadi1991semantics}, and they undoubtedly share some sort of similarity at a high level, the concept of event is however different.
In fact, its formalism has been built on top of a criticism to a lack of formality in the BAN logic~\cite{woo1993semantic}; in particular, an event may record something that cannot be interpreted as a {\em belief}.
The generic proposition $e\of{a_1, \dots, a_n}$ is used as a short notation to say that there exists (at least) a trace which recorded such event.

The reasoning engine of ProVerif will execute a {\em main} process and record traces of execution.
At the same time, the attacker's knowledge and the tables, if any, are accordingly updated.
Security properties are eventually checked by inspecting traces, tables, the attacker's knowledge, and, for equivalences, relations between traces and processes.
We refer to the paper by Blanchet~\cite{blanchet2001efficient} for additional details.

\subsection{Modelling the SPEKE protocol}
\label{sec:formal-model-protocols}
We formally model the following variants of the SPEKE protocol in the Applied Pi Calculus~\cite{ryan2011applied}: the original Jablon's protocol~\cite{jablon1996strong},
the ones in IEEE P1363.2:D26~\cite{ieeep1363.2d26} and ISO/IEC 11770-4:2006~\cite{isoiec11770-4:2006}, the earlier patch proposed by Hao and Shahandashti in 2014~\cite{hao2014speke}, and the final patch described in this paper and included into ISO/IEC 11770-4:2017~\cite{isoiec11770p4}, each in two modes:
\begin{itemize}
  \item without explicit key confirmation,
  \item with explicit key confirmation as described in the respective documents.
\end{itemize}

It is worth noting that a meaningful key exchange process should always be completed with some form of key confirmation, let it be explicit or implicit. The explicit key confirmation is  realized by executing the explicit key confirmation procedure, which requires extra rounds of communication. But the explicit key confirmation procedure is optional~\cite{ieeep1363.2d26,isoiec11770-4:2006}: without it, the key confirmation is deferred to the secure communication stage, and this is called \emph{implicit} key confirmation~\cite{stinson2005cryptography}. However, the exact mechanisms for \emph{implicit} key confirmation are not specified in~\cite{jablon1996strong,ieeep1363.2d26,isoiec11770-4:2006}, which makes it difficult to model SPEKE with implicit key confirmation. To address this issue, we assume the implicit key confirmation is realized in the secure communication stage by prepending the first encrypted message with an explicit key confirmation string as defined in the respective explicit key confirmation procedure. Thus, our formal model treats SPEKE with implicit and explicit key confirmations as essentially the same with the only difference being that the latter requires additional rounds of communication. 

In the model, we formally specify the following:

\textbf{The two parties}.
All variants of the SPEKE protocol involve two parties: the Initiator $I$ and the Responder $R$. They are modelled as two processes $P_I$ and $P_R$. We use the initiator and the responder for the convenience of naming in our model. Essentially we assume that one party initiates the protocol by sending data in the first flow, and the other party responds by sending data in the second flow. Thus a one-round protocol is implemented in two flows. This does not change the security analysis of the protocol. Below we give the ``vanilla'' specification of the protocol. In the ``vanilla'' specification, we abstract out key reconstruction by a function symbol $kdf$, and the confirmation messages sent by the Initiator and the Responder are abstracted by the symbols $kcf_I$ and $kcf_R$ respectively. The actual specification of each variant has its own definitions of $kdf$, $kcf_I$ and $kcf_R$ to capture the differences between the variants.   

\begin{figure}[htbp]
\small
\caption{The processes for the Initiator, $P_I$, and the Responder, $P_R$. In the above specification, the notation $= X$ means abort if the incoming value is not $X$.}
\label{fig:processes_initiator_responder}
\begin{minipage}{\columnwidth}
\begin{tikzpicture}
  \node[anchor=south west,inner sep=0] at (0,0) {
  \begin{minipage}[c]{0.04\textwidth}
    {\fontsize{7pt}{\baselineskip}\color{gray!90!black}
    \vskip 1em
    \begin{align*}
      1 \\ 2 \\ 3 \\ 4 \\ 5 \\ 6 \\ 7 \\ 8 \\ 9
    \end{align*}}%
  \end{minipage}%
  {\color{gray}\vrule height 7em}%
  \begin{minipage}[c]{0.48\textwidth}
    \begin{equation*}
      \begin{split}
        P_I \gets\ & \piread{c}{\pair{I}{R}}; \\
                  & \piget{t}{=I, =R, g}{}{}{} \\
                  & \nu {\color{red}x}. \pilet{X = g^{\color{red}x}}{} \\
                  & \piwrite{c}{\pair{I}{X}}; \\
                  & \piread{c}{\pair{=R}{Y}}; \\
                  & \pilet{k = kdf}{} \\
                  & \piwrite{c}{kcf_I}; \\
                  & \piread{c}{=kcf_R}; \\
                  & \piwrite{c}{\mn{enc}\of{k, m}}; \\
      \end{split}
    \end{equation*}%
  \end{minipage}%
  \begin{minipage}[c]{0.48\textwidth}
  \begin{equation*}
    \begin{split}
      P_R \gets\ & \piread{c}{\pair{I}{R}}; \\
                & \piget{t}{=R, =I, g}{}{}{} \\
                & \nu {\color{red}y}. \pilet{Y = g^{\color{red}y}}{} \\
                & \piwrite{c}{\pair{R}{Y}}; \\
                & \piread{c}{\pair{=I}{X}}; \\
                & \pilet{k = kdf}{} \\
                & \piread{c}{=kcf_I}; \\
                & \piwrite{c}{kcf_R}; \\
                & \piwrite{c}{\mn{enc}\of{k, m}}; \\
    \end{split}
  \end{equation*}%
  \end{minipage}
  };
  \fill[draw=cyan!70!black,fill=cyan!10!blue!20!white,fill opacity=0.05] (1.11,0.9) rectangle (3.85,3.63);
  \fill[draw=cyan!70!black,fill=cyan!10!blue!20!white,fill opacity=0.05] (5.2,0.9) rectangle (7.9,3.63);
\end{tikzpicture}
\end{minipage}
\end{figure}

As can be easily seen in Figure~\ref{fig:processes_initiator_responder}, the code inside the boxes is the part modelling the protocol scheme depicted in 
Figure~\ref{fig:speke_jablon}, where the key reconstruction part is abstracted by the function symbol $kdf$, and the confirmation messages sent by the Initiator and the Responder are abstracted by the symbols $kcf_I$ and $kcf_R$ respectively, where we omit their arguments for simplicity. We highlight the symmetric nature of the protocol letting both processes to write to the channel simultaneously.

The other lines (outside the box) in Figure~\ref{fig:processes_initiator_responder} serve to model the behaviour of the protocol and to verify security properties.
In particular, the first line is to let the processes know the identities involved in the protocol; they read them from the channel $c$ at the very beginning.
The second line checks whether the password table contains a suitable password to communicate to the other party; otherwise, they abort.
The last line is useful to verify the secrecy of the shared key $k$ through the privacy of the message $m$, and the perfect 
forward secrecy.
The details are deferred to Section~\ref{sec:security-properties} where we discuss the security properties.



\textbf{The pre-shared password}.
A table $t$ of passwords is filled with all secret group generators that would be calculated from the passwords, i.e., $g \in \typeset{G}$ is the secret group generator for $A$ and $B$.
From the point of view of the symbolic protocol design, sharing a password and then computing the generator is the same as having directly shared the secret generator.

\textbf{The main process}.
Informally, the main process $P$ is an infinite repetition of the parallel execution of the Initiator's process $P_I$ and the Responder's process $P_R$.
Due to the symmetric nature of the SPEKE protocol, the naive implementation of the main process brings {\em false} attacks where the Initiator speaks to itself.
To avoid this issue, we must explicitly support the session within the two parties.
However, the session $s$ is not private information, and we disclose it to the attacker by outputting it to the insecure channel $c$, i.e. $\piwrite{c}{s}$.
At this point, we have the infinite repetition of the following process: $!\pbrackets{\nu s. \piwrite{c}{s}; \pbrackets{P_I | P_R} }$.
The two parties would never engage the protocol if they do not share the password. For this reason, we have an environment process $P_P$ which is in charge of inserting shared passwords into a table that can be accessed by $P_I$ and $P_R$, but not the attacker. In order to record events and verify correspondences, we also have a process $P_A$, which records the agreements between the parties through events.

%

Finally, the main process $P$ that the tool checks has the following structure:
\begin{equation*}
  P \gets \pbrackets{P_P | \pbrackets{!\pbrackets{\nu s. \piwrite{c}{s}; \pbrackets{P_I | P_R} }} | !P_A  }.
\end{equation*}

The process $P_A$ collects information from two tables, one filled in by the Initiator and the other by the Responder.
We emphasise that the protocol can be initiated by either of them and the two tables are put together recording a single event. For security properties that do not require tables to record events, the process will be simply $\procstop$; otherwise, depending on the property to prove, the events $e_S$ end $e_C$ can be recorded, where $e_S$ means that the involved parties in the protocol agree with the participants, the session, and the reconstructed session key at the end of the protocol, and $e_C$ means that the involved parties in the protocol agree with the participants, the session, the secret group, the secret nonce, and the reconstructed session key at the end of the protocol.

\subsection{Security properties}
\label{sec:security-properties}
The security properties are modelled as follows.
\subsubsection{Correctness}
This property checks whether the protocol gives authentication and key distribution in presence of honest parties~\cite{woo1993semantic}.
Even though this property is generally the easiest to prove, it should not be neglected when formally modelling a protocol, in order to avoid either logical or typographic errors.
To check the correctness of the models, we need to reconstruct the session key $kdf$.
Its implementation depends on the SPEKE variants.


Formally, for all the sessions and nonce exponents, we require that there exists at least a trace in which the event collecting private and shared values of the participants is recorded and is such that the two honest participants agree on their identities, the password, and the session key with the right formula.
\begin{align*}
  \forall s & \in \typeset{S}, x, y \in \intset\kstar_q, g \in \intset\kstar_p.\; \\
    & \begin{array}{r@{}l}
        e_C ( & A, B, s, g, x, kdf\of{A, B, g^x, g^y, g^{xy}, s}, \\
              & A, B, s, g, y, kdf\of{A, B, g^x, g^y, g^{xy}, s} )
      \end{array}
\end{align*}
where $A$ and $B$ are honest parties and $g$ is the generator calculated from the shared password.
If such an event is raised, then there exists a run of the protocol in which the two parties have authenticated each other and they have correctly computed the session key.

\subsubsection{Secrecy of the pre-shared password}

We proved the secrecy of the password through observational equivalence.
Formally, if we call $\pi_g$ the process describing the protocol where two honest parties $A$ and $B$ share the password $g$, and $\pi_{g^{\prime}}$ the same protocol but with $g^{\prime}$ instead of $g$, then the observational equivalence $\pi_g \approx \pi_{g^{\prime}}$ describes the property that any attacker cannot distinguish between the two runs of the protocol with probability (non-negligibly) better than a blind guess, and therefore no extra information about the secret password can be gained.

\subsubsection{Implicit key authentication}
Implicit key authentication is verified when only the two participants can reconstruct the session key.
This concept is modelled by using the key to encrypt a secret message with deterministic encryption.
We then check for observational equivalence of two runs of the processes $P_I$ and $P_R$ where in the last line (Figure~\ref{fig:processes_initiator_responder}) the message encrypted is provided by a choice, $\piwrite{c}{\mn{enc}\of{k, \sbrackets{m,m^{\prime}}}}$.
Similar to how we determine the secrecy of the password, if we call $\pi_m$ the process describing the protocol where two honest parties $A$ and $B$ encrypt $m$, and $\pi_{m^{\prime}}$ the same protocol but with $m^{\prime} \neq m$, then the observational equivalence $\pi_m \approx \pi_{m^{\prime}}$ is verified.
If the observational equivalence holds and therefore the message $m$ remains secret, it trivially follows that the shared key is at least as secret as $m$.
In fact, the decryption function is public, and the reconstruction of the key will irredeemably compromise the secrecy of $m$.



\subsubsection{Explicit key authentication}
Explicit key authentication is verified when only the two participants can reconstruct the session key, and they actually do.
It is therefore defined as implicit key authentication {\em and} an agreement on the computed key for the same session.
Formally,
\begin{align*}
  \forall h_1, h_2 & \in \typeset{H}, s \in \typeset{S}, k, k^{\prime} \in \typeset{K}.\; \\
    & e_S\of{h_1, h_2, s, k, h_1, h_2, s, k^{\prime}} => k = k^{\prime}
\end{align*}
In other words, in a trace of execution the presence of the event $e_S$ where the first and fifth  arguments being equal (agreement on the initiator), the second and the sixth being equal (agreement on the responder), and the third and the seventh being equal (agreement on the session) implies that the fourth and the eighth are equal (equivalence of the reconstructed key).
When this property is true, a protocol completed between two authenticated parties in the same session guarantees that the parties agree on the session key.
This property, along with the implicit key authentication, gives explicit key authentication.

\subsubsection{Weak and strong entity authentication}
Weak entity authentication guarantees that two parties are indeed speaking to each other.
Strong entity authentication requires agreement on other values than the mere entities.
These values are supposed not to be injected, produced or inferred by an attacker.
Those two properties share similarities in their formality.
The events involved are 1) $e_{I}$ to record that the initiator $I$ believes that it has started a protocol with the responder $R$; 2) $e_{R}$ to record that $R$ believes that it has started a protocol with $I$; 3) $e_{IR}$ to record that $I$ believes that it speaks to $R$ at the end of the protocol, and 4) $e_{RI}$ to record that $R$ believes that it speaks to $I$ at the end of the protocol.
The first and the third are recorded by the honest initiator, while the the second and the fourth by the honest responder.
Mutual {\em weak} authentication is provided by the two following symmetric correspondences, one for each honest party:
\begin{align*}
  \forall h_1, h_2 \in \typeset{H}.\; & e_{IR}\of{h_1, h_2} => e_{R}\of{h_1, h_2} \\
  \forall h_1, h_2 \in \typeset{H}.\; & e_{RI}\of{h_1, h_2} => e_{I}\of{h_1, h_2}
\end{align*}
And mutual strong authentication by the following:
\begin{align*}
  \forall h_1, h_2 & \in \typeset{H}, s \in \typeset{S}, k \in \typeset{K}.\; \\ & e_{IR}\of{h_1, h_2, s, k} => e_{R}\of{h_1, h_2, s, k} \\
  \forall h_1, h_2 & \in \typeset{H}, s \in \typeset{S}, k \in \typeset{K}.\; \\ & e_{RI}\of{h_1, h_2, s, k} => e_{I}\of{h_1, h_2, s, k}
\end{align*}
where they agree also on the session and the exchanged session key.
Agreeing on the session will prevent any replay attack from other sessions, even concurrent, while agreeing on the key will guarantee that no attacker can let two authenticated parties not to share the same key.
However, a key-malleability attack is still possible even if the protocol can achieve strong entity authentication.

\subsubsection{Perfect forward secrecy}
Usually, key exchange protocols verify (or claim) the perfect forward secrecy (PFS) property.
For the password authenticated key exchange protocols, this property means that if passwords are compromised, the {\em past} session keys derived from such passwords still remain secret.
Hence, an adversary can only keep a record of past communication which has not been compromised.
We can reformulate this concept as a {\em passive} adversary whom is given the password and eavesdrops (unbounded number of) executions of the protocol trying to reconstruct any of the session keys.
In practice, to verify this property we disclose the secret generator $g$ to the attacker, $\piwrite{c}{g}$, then we query the non-interference property on the encrypted message.
Since the passive attacker can compute any decryption, the non-interference property captures the perfect forward secrecy, i.e., if the encrypted message cannot be reconstructed, it must be that any session key cannot be reconstructed either.

\subsubsection{Bilateral UKS}
Informally, a successful bilateral UKS attack makes two honest parties $I$ and $R$ believe that they share $k$ with some other party~\cite{chen2008bilateral}.
To capture this attack, we use the following correspondence:
\begin{align*}
  \forall h_1, & h_2, h_1^{\prime}, h_2^{\prime} \in \typeset{H}, s, s^{\prime} \in \typeset{S}, k \in \typeset{K}.\; \\
  & e_S\of{h_1, h_2, s, k, h_1^{\prime}, h_2^{\prime}, s^{\prime}, k} => h_1 = h_1^{\prime} \land h_2 = h_2^{\prime}
\end{align*}
If an initiator and a responder recorded the same key, then it must be that they agree on the entities.
If we required that they should agree on the session too, then we could put $s = s^{\prime}$ in logical AND with the two equivalences.
On the contrary, if we wanted to force the tool to show bilateral UKS attacks in the same session, we could state $s = s^{\prime}$ as a premise.




\subsubsection{Impersonation attack}
The impersonation attack is a problem that generally affects SPEKE protocols and an instance of such an attack has been shown in Section~\ref{sec:impersonation-attack}. To formalise this attack, we build a model in which there exists only one honest party and the attacker. In this case, if the honest party ever shares a key with another party, then the other party must be the attacker, and the attacker must impersonate another honest party in order to run the protocol up to this point. In fact, all SPEKE variants without key confirmation phase are vulnerable to this attack.


To verify, we can check for every honest party, session and key, the event of authenticating the other party is not recorded in any trace (i.e. the adversary cannot establish a shared key with the honest party). Formally, we check the following property:
\begin{align*}
  \forall h_1, h_2 \in \typeset{H}, & s \in \typeset{S}, k \in \typeset{K}.\; \\
    & \neg \pbrackets{e_{RI}\of{h_1, h_2, s, k} \lor e_{IR}\of{h_1, h_2, s, k}}.
\end{align*}

\subsubsection{Sessions swap}
A man-in-the-middle is able to perform the sessions swap attack if it can let a honest party in some session $s$ share a key with another honest party in some other concurrent session $s^{\prime}$ and vice versa.
This attack occurs in a key-exchange protocol where the key does not depend on the session.
Formally, we say that for every two parties and for every key, the presence of an agreement on the Initiator, Responder, and session key must imply the equivalence of the sessions.
\begin{align*}
  \forall h_1, h_2 \in \typeset{H}, & s, s^{\prime} \in \typeset{S}, k \in \typeset{K}.\; \\
    & e_S\of{h_1, h_2, s, k, h_1, h_2, s^{\prime}, k} => s = s^{\prime}.
\end{align*}

\subsubsection{Malleability}
The malleability of the session key is an attack that affects many variants of the SPEKE protocols, and it was described in Section~\ref{sec:malleability-attack}.
Capturing the malleability attack in ProVerif requires more efforts than other attacks, because it is based on an {\em extra} level of group exponentiation equality (three commutative exponents). This results in a larger search space when the reasoning engine checks the property, and ProVerif slows down considerably (taking minutes instead of milliseconds to verify the non-malleability property on a 3.2 GHz computer with 64 GB RAM running Linux). To detect malleability, we require the two honest parties to write into a table some values they agree with, plus their secret fresh exponents and the secret generator (the password).
This way, when checking for correspondence, we can check whether the key is indeed what is expected with regard to the private inputs of the parties.


Formally, for every pair of parties, session, generator, two exponents and key, where the parties agree on the identities, the session, the generator and the key (they cannot agree on the other party's secret), then the key they agree on is computed equivalently to the formula provided by the protocol.
\begin{align*}
  \forall & h_1, h_2 \in \typeset{H}, x, y \in \intset\kstar_q, g \in \intset\kstar_p, s \in \typeset{S}, k \in \typeset{K}.\; \\
  & e_C\of{h_1, h_2, s, h, x, k, h_1, h_2, s, g, y, k} => \\
  & k = kdf\of{a, b, g^x, g^y, g^{xy}, s} \lor kdf\of{b, a, g^y, g^x, g^{xy}, s}.
\end{align*}
Note the key $k$ may have two different values depending on in the protocol who is the initiator and who is the responder.


\begin{table*}
\caption{Summary of results on efficiency and formal verification in ProVerif}
\label{tbl:speke-security}
\vspace{-3mm}
\begin{center}
\begin{tabular}{c@{ }c@{ }c@{ }c@{ }c@{ }c@{ }c@{ }c@{ }c@{ }c@{ }c@{ }c@{ }c}
& \textbf{Variants}                                                                                                                    & \textbf{RND} & \textbf{RND-E} & \textbf{IKA} & \textbf{EKA} & \textbf{WA} & \textbf{SA} & \textbf{IMP} & \textbf{SS} & \textbf{PFS} & \textbf{UKS} & \textbf{MAL} \\ \hline\toprule
 & \multicolumn{1}{c}{Jablon 1996 paper~\cite{jablon1996strong}} & 1 & 3 & $\resultve$  & $\resultfa$  & $\resultfa$ & $\resultfa$ & $\resultfa$  & $\resultfa$ & $\resultve$   & $\resultfa$  & $\resultfa$  \\
& IEEE P1363.2:D26~\cite{ieeep1363.2d26} & 1 & 3 & $\resultve$  & $\resultfa$  & $\resultfa$ & $\resultfa$ & $\resultfa$  & $\resultfa$ & $\resultve$   & $\resultfa$  & $\resultve$  \\
& ISO/IEC 11770-4:2006~\cite{isoiec11770-4:2006} & 1 & 3 & $\resultve$  & $\resultfa$  & $\resultfa$ & $\resultfa$ & $\resultfa$  & $\resultfa$ & $\resultve$   & $\resultfa$  & $\resultve$  \\
& Hao and Shahandashti\cite{hao2014speke} & 1 & 2 & $\resultve$  & $\resultve$  & $\resultve$ & $\resultve$ & $\resultve$  & $\resultve$ & $\resultve$   & $\resultve$  & $\resultve$  \\
& P-SPEKE (ISO/IEC 11770-4:2017) & 1 & 2 & $\resultve$  & $\resultve$  & $\resultve$ & $\resultve$ & $\resultve$  & $\resultve$ & $\resultve$   & $\resultve$  & $\resultve$  \\
\bottomrule
\end{tabular}
\end{center}
The results are grouped by variants with and without key confirmation phase (KC).
\newline
\textbf{\em Legend}. \textbf{Round efficiency}: without explicit key confirmation (RND), with explicit key confirmation (RND-E). \textbf{Security properties}: Implicit Key Authentication (IKA), Explicit Key Authentication (EKA), Weak Entity Authentication (WA), Strong Entity Authentication (SA), Impersonation resilience (IMP), Sessions Swap resilience (SS), Perfect Forward Secrecy (PFS), bilateral Unknown Key-Share resilience (UKS), and Malleability resilience (MAL).
\textbf{Outcomes}: ($\resultve$) - verified, ($\resultfa$) - attacks found, ($\resultna$) - not applicable.
\end{table*}

\section{Summary of results}
\label{sec:results}
The ProVerif scripts that we created to model and verify the protocols are available at GitHub~\cite{githubsource}. 
There are in total 54 scripts related to this paper, each for a different variant and a property. ProVerif will give one of the following four responses: (i) the property is true, (ii) the property is false, (iii) the property cannot be proved, and (iv) non-termination.
The results are summarised in Table \ref{tbl:speke-security}. 
The proposed patch (as well as the patch in~\cite{hao2014speke}) improves the round efficiency over the previous SPEKE variants~\cite{jablon1996strong,ieeep1363.2d26,isoiec11770-4:2006} by allowing the explicit key confirmation steps to be completed within one round. As a result, it requires only 2 rounds to finish the key exchange with explicit key confirmation as opposed to 3 rounds previously. All variants have the Implicit Key Authentication (IKA) property, confirming that the session key will not be learned by the attacker, and that the attacker cannot get confidential information by eavesdropping. This does not contradict the impersonation attack shown in Section~\ref{sec:new-attacks}, since that attack works without the adversary learning the session key. However, that attack demonstrates that the adversary is able to manipulate the two parallel sessions to make them generate \emph{identical} session keys. Consequently, the adversary is able to pass the explicit key confirmation by replaying messages. This is confirmed by our formal analysis that the original SPEKE~\cite{jablon1996strong}, and the SPEKE in standards~\cite{ieeep1363.2d26,isoiec11770-4:2006} do not fulfil the explicit key authentication property. Also, the 
existence of the impersonation attack shows that these variants do not fulfil the weak/strong entity authentication which concerns assuring the identities of the entities involved in the key exchange protocol. The proposed patch prevents the Session Swap attack (SS), the UKS attack, and the Malleability (MAL) attack by making the session key depend on the session, the identities, and the transcript of the key exchange process. We emphasise that these security properties are verified {\em before} any key confirmation either implicit or explicit. To guarantee that the participants are mutually authenticated, the key confirmation becomes necessary. Such key confirmation must include all of the key points above, i.e., session, identities, and a transcript of the key exchange messages, so avoiding the above mentioned attacks. Including only the identities allows to verify {\em weak} entity authentication only.

Our formal analysis using ProVerif confirms that our proposed patch prevents the two attacks as identified earlier. However, this analysis does not constitute a complete proof of security for SPEKE, as one might expect from formal authenticated key exchange models~\cite{bellare1993entity,bellare2000authenticated,katz2001efficient,goldreich2001session,abdalla2015security}. In particular, we have not proved that SPEKE is resistant against off-line dictionary attacks based on standard security assumptions such as DDH or CDH. We highlight that the original SPEKE was designed without a security proof. Retrospective efforts to prove the security of a protocol based on standard number theoretical assumptions may turn out to be very hard if not impossible. We leave further analysis of SPEKE to future work.

\section{Conclusions}
\label{sec:conclusions}
The SPEKE protocol was firstly proposed by Jablon over two decades ago. Since then, it has been adopted by international standards, and built into smart phones and other products. We identified two weaknesses in the standardized SPEKE specification, which affect all implementations that follow the IEEE 1362.3 and ISO/IEC standards. Accordingly we proposed a patched SPEKE to address the identified issues. We formally modelled the discovered attacks against SPEKE and proved that the proposed patch was immune to these attacks. In addition, we contributed to improve the round efficiency of the protocol in the key confirmation phrase. Our proposed patch and the improved key confirmation procedure have been included into the latest revision ISO/IEC 11770-4 published in July 2017. However, the SPEKE specification in IEEE P1363.2 (which is currently not maintained) remains unfixed. 

The problems in SPEKE identified in this paper have evaded 20 years cryptanalysis (informal and formal) by the security and standardization communities. The initial discovery of the two attacks on SPEKE was down to manual analysis, which was later formally verified by applying the ProVerif tool. The mechanised proofs that we produce are not only helpful for proving security properties of similar protocols, but also for preventing the same problems in the future. This shows that traditional human cryptanalysis, in conjunction with modern automated proof techniques, is useful in improving security protocols, especially those that have been included in international standards.



\section*{Acknowledgements}

We thank Professor Liqun Chen for her invaluable advice and comments on revising SPEKE in ISO/IEC 11770-4.

\bibliographystyle{abbrv}
\bibliography{SPEKE-bib}

\end{document}